\documentclass[twocolumn,showpacs,preprintnumbers,amsmath,amssymb]{revtex4}
\usepackage{graphicx}% Include figure files
\usepackage{dcolumn}% Align table columns on decimal point
\usepackage{bm}% bold math

\begin{document}

\preprint{APS/123-QED}

\title{Glassy dynamics in thin films of polystyrene}
% Force line breaks with \\

\author{Koji Fukao}%\cite{A}}
 \email{kfukao@se.ritsumei.ac.jp}
 \altaffiliation[]{Corresponding author. 
}
\affiliation{%
Department of Physics, Ritsumeikan University, Noji-higashi 1-1-1, 
Kusatsu, 525-8577, Japan
}%
\author{Hiroki Koizumi}%
\affiliation{%
Department of Polymer Science, Kyoto Institute of Technology, 
Matsugasaki, Kyoto 606-8585, Japan
}%

\date{\today}% It is always \today, today,

\begin{abstract}
Glassy dynamics was investigated for thin films of atactic polystyrene 
by complex electric capacitance measurements using dielectric
relaxation spectroscopy. During the isothermal aging process the real part 
of the electric capacitance increased with time, whereas the 
imaginary part decreased with time.
It follows that the aging time dependences of real and imaginary parts
 of the electric capacitance were primarily associated with change 
in volume (film thickness) and
 dielectric permittivity, respectively. 
Further, dielectric permittivity showed memory and rejuvenation effects in a
similar manner to those  observed for poly(methyl methacrylate) thin
 films. On the other hand, 
volume did not show a strong rejuvenation effect.
\end{abstract}

%71.55.Jv  Disordered structures; amorphous and glassy solids
%81.05.Lg  Polymers and plastics; rubber; synthetic and natural fibers;
%          organometallic and organic materials
%77.22.Ch  Permittivity (dielectric function)

\maketitle

\section{Introduction}
Amorphous materials show a glass transition as the temperature 
decreases in the liquid state. Throughout the glass transition, 
the molecular mobility due to the $\alpha$-process is frozen and 
the material transforms into a glassy state.
Polymeric glass shows structural changes during the aging process, 
even below the glass transition temperature $T_{\rm g}$. Corresponding 
 physical changes are observed with these structural 
changes~\cite{Struick,Kovacs,Bouchaud}. 
These changes are known as physical aging, and are related to the glassy
dynamics.  
This is regarded as an important property characteristic of
disordered materials including; polymer
glasses~\cite{Bellon1,Fukao2,Fukao5}, spin
glasses~\cite{Lefloch,Vincent} and other disordered 
systems~\cite{Doussineau,Leheny,Kircher}.

In our previous papers~\cite{Fukao2,Fukao5}, the glassy
states of poly(methyl
methacrylate) (PMMA) show a decrease in both the real and imaginary 
parts of the
complex dielectric constant with increasing time during the
isothermal aging process.  The glassy states show 
{\it memory and rejuvenation 
phenomena} for thermal treatment of a constant rate mode 
and of a temperature cycling mode, as shown below. In the constant rate mode,
the temperature changes from high temperature 
above $T_g$ to an aging temperature, at which the sample is
aged isothermally and the dielectric susceptibility decreases
with increasing aging time during the process.
This decrease in dielectric susceptibility is associated with 
the relaxation of the system into the equilibrium states. The change
to the equilibrium is called {\it aging}. Then, the temperature 
subsequently decreases from the isothermal aging temperature 
to room temperature at a
constant rate. A comparison of the value of the dielectric susceptibility
at room temperature after the isothermal aging to 
that without isothermal aging, revealed that
the values were in good agreement. This suggests that 
the decrease in dielectric susceptibility induced by
the isothermal aging is totally compensated during the cooling process 
from the isothermal aging temperature to room temperature. This change in
dielectric susceptibility has the opposite direction compared to the
change due to the aging. Therefore, the change in dielectric
susceptibility induced during the cooling process after the isothermal
aging is called {\it rejuvenation}, and hence the system 
is {\it rejuvenated}
as far as the dielectric response is concerned. However, during 
the subsequent heating process from room temperature,
the dielectric susceptibility deviates from the value observed during
the cooling process without isothermal aging, and shows a maximum
deviation just above the aging temperature as if the sample remembered 
the isothermal aging  it experienced during the preceding
cooling process. This behavior is called {\it memory effect} and 
it suggests that the thermal history is maintained as a {\it memory}.

In the case of amorphous polymers, it is well known that the density of
the polymers increases due to the isothermal aging below $T_{\rm g}$,
and this density change is maintained even at room
temperature~\cite{Struick}.  On the other hand, there is
 {\it no rejuvenation} for the density. This finding appears to be 
inconsistent with the fact that the dielectric as-susceptibility shows
rejuvenation.
Herein, we investigate the aging behavior of thin films of
atactic polystyrene (a-PS) using electric
capacitance measurements in order to elucidate the nature of the 
glassy dynamics in polymer glasses.

\section{Experimental}

Thin films of atactic polystyrene (a-PS) with two different thicknesses, 
14 nm and 284 nm, were prepared by spin-coating from a
toluene solution of a-PS on an Aluminum (Al)-deposited glass substrate. 
The thickness was controlled by changing the concentration of solution. 
The samples of a-PS used in this study were purchased from 
Aldrich Co., Ltd. ($M_{\rm w}$~=~$1.8\times 10^6$, 
$M_{\rm w}/M_{\rm n}$~=~$1.03$). 
After annealing at 70 $^{\circ}$C under vacuum for two days 
to remove solvents, Al was again vacuum-deposited to 
serve as an upper electrode. Several heating cycles through and above
a bulk value of $T_{\rm g}$ were carried out before the capacitance  
measurements were taken for the relaxed as-spun films in order to obtain 
reproducible results. 
The values of $T_{\rm g}$ in the thin films with $d$~=~14~nm and 284~nm were
350~K and 368~K, respectively.
The thickness was evaluated from the value of the electric capacitance at 
293~K for the as prepared films before measurements in the same manner 
previously reported ~\cite{Fukao1,Fukao3}. %,Fukao4}. 

The $T_{\rm g}$ of the two thin films differs by about 18
K. According to the recent report on $T_{\rm g}$ of thin films of 
polystyrene, the effect of oxidation can be a possible origin of  
the reduction of $T_g$~\cite{Serghei1}. However, we performed 
$T_{\rm g}$ measurements for each thin film several times and 
confirmed that the reduction of $T_{\rm g}$ is not due to the
oxidation effect in the present case.

Capacitance measurements were carried out using an LCR meter (HP4284A) for the 
frequency $f$ =~1~kHz during the cooling and heating  
processes between 380~K and 273~K at a rate of 1~K/min 
as well as during isothermal aging at various aging temperatures
$T_a$ (=315~K $\sim$ 351~K).
The complex electric capacitance of the sample
condenser $C^*(\equiv C'-iC'')$ was measured as a function of 
temperature $T$ and aging time $t_{\rm w}$. 
The value of $C^*$ was converted into 
the dynamic (complex) dielectric constant 
$\epsilon^*(\equiv \epsilon'-i\epsilon'')$ by 
dividing the $C^*(T)$ by the geometrical capacitance $C_0(T_0)$ 
at a standard temperature $T_0$.
The value of  $C^*$ is given by the relation 
$C^*=\epsilon^*\epsilon_0\frac{S}{d}$ and
$C_0=\epsilon_0 \frac{S}{d}$, 
where $\epsilon_0$ is the permittivity in vacuum, $S$ is the area of 
the electrode and $d$ is film thickness. 
For evaluation of $\epsilon^*$ and $C_0$, 
we use the thickness $d$ which is determined at $T_0=$~293~K and 
$S=$~8$\times$10$^{-6}$ m$^2$.

\section{Results and Discussion}

\subsection{Aging dynamics}

\begin{figure}%[b]
\includegraphics*[width=7.2cm]{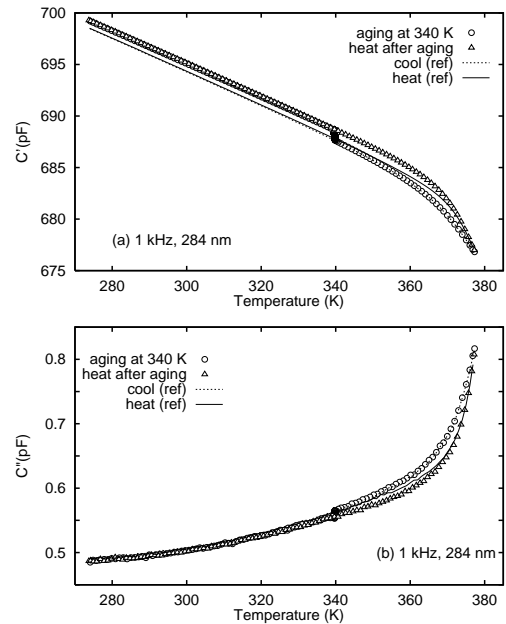}% Here is how to import EPS art
\caption{\label{fig:1} Temperature change in (a) the real part of electric
 capacitance $C'$ and (b) the imaginary part $C''$ for 
 a-PS films with thickness of 284~nm for frequency 1~kHz. Temperature changes
between 380~K and 273~K including isothermal aging at 340~K.
% (See the detailed explanation in text).
}
\end{figure}

Figure 1 shows the change in the real and imaginary parts of the complex
electric capacitance for a-PS films with $d$~=~284~nm and
$f$=~1~kHz. The temperature of the sample was changed as follows: After 
several temperature cycles in order to stabilize the
measurements, the temperature changed from 380~K to 273~K (cooling
process, dotted line) and then from 273~K to 380~K (heating process, solid
line). The data obtained for the cooling and heating processes are
used as a reference data. The sample was cooled from 380~K
to 340~K($\equiv T_a$), the temperature was maintained at $T_a$ for
20 hours, and  then the sample was again cooled down to 273 K after
isothermal aging (open circles). Finally, the sample was again heated 
from 273~K to 380~K (open triangles). 

In Fig.~1(a), the reference data shows that the real part of
the electric capacitance $C'$ is a
decreasing function of temperature and the slope of $C'$ with respect to 
temperature gradually changed around 365~K. Thus, there is a
difference between the heating and cooling processes. This gradual change 
corresponds to the glass transition. 
Furthermore, from the data 
for the subsequent cooling process including isothermal aging, 
{\it $C'$ was observed to increase with aging time} during isothermal aging
at $T_a$. In this case, the value of $C'$ increased by 0.3 \% 
for isothermal aging at 340~K for 20 hours. 

Contrary to $C'$, the imaginary part of the electric capacitance $C''$ 
was an increasing function of temperature during the cooling and heating
processes at a constant rate. Above 380~K a decrease in the peak was 
expected due to the $\alpha$-process.
Further, during the isothermal aging process at 340~K, $C''$ was
observed to decrease with
increasing aging time. The change in $C''$ for aging at 340~K for
20 hours was about 0.2 pF, which corresponds to a decrease by 3\%.
This fraction was an order of magnitude larger than the increase in $C'$ 
for the same aging process. 

\begin{figure}%[b]
%\includegraphics*[width=7.2cm]{./graph_ps025a_5f1_a.eps}% Here is how to
 % import EPS art

%\includegraphics*[width=7.2cm]{./graph_ps025a_5f1_b.eps}% Here is how to import EPS art
\includegraphics*[width=7.2cm]{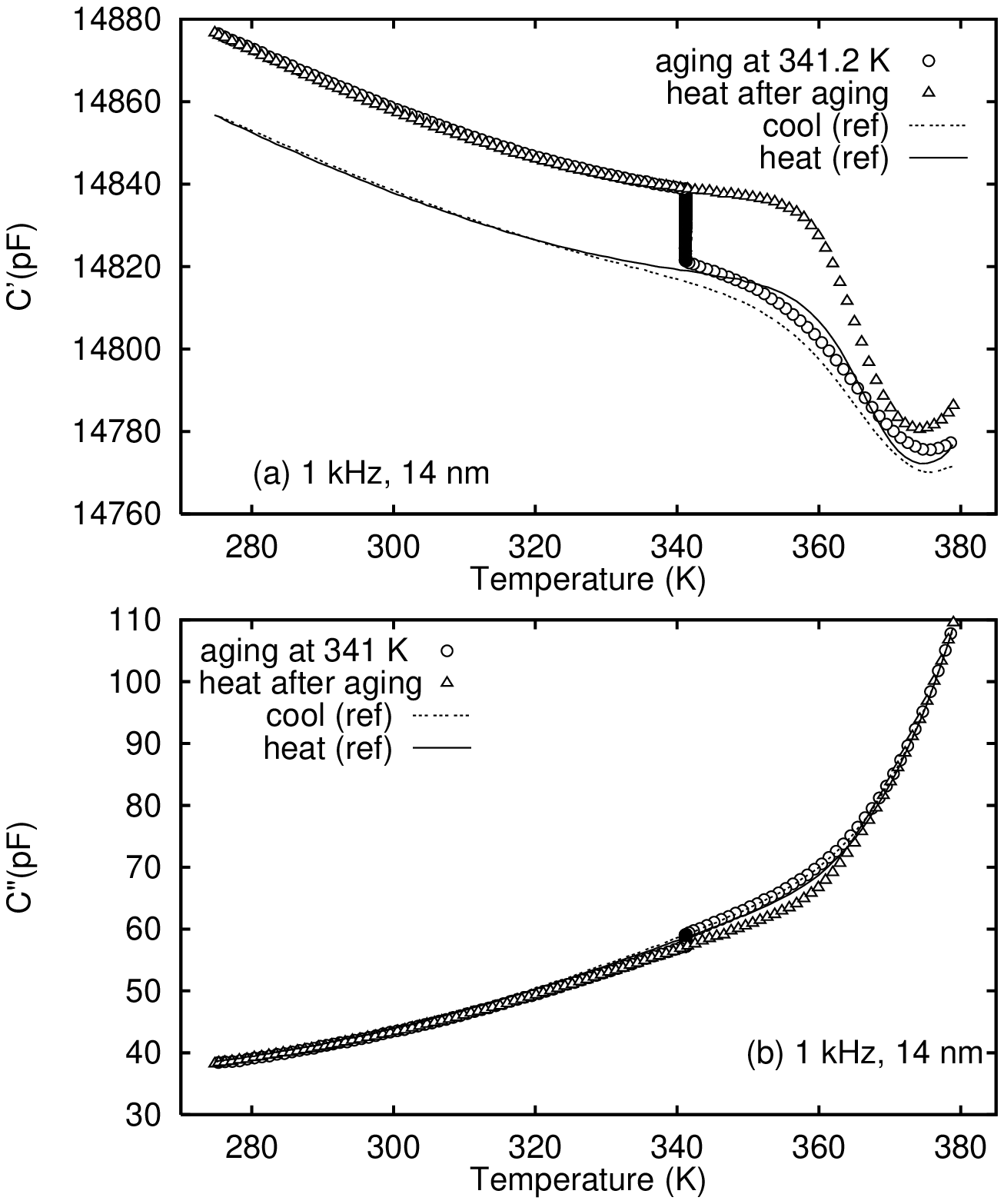}% Here is how to
 % import EPS art

\includegraphics*[width=7.2cm]{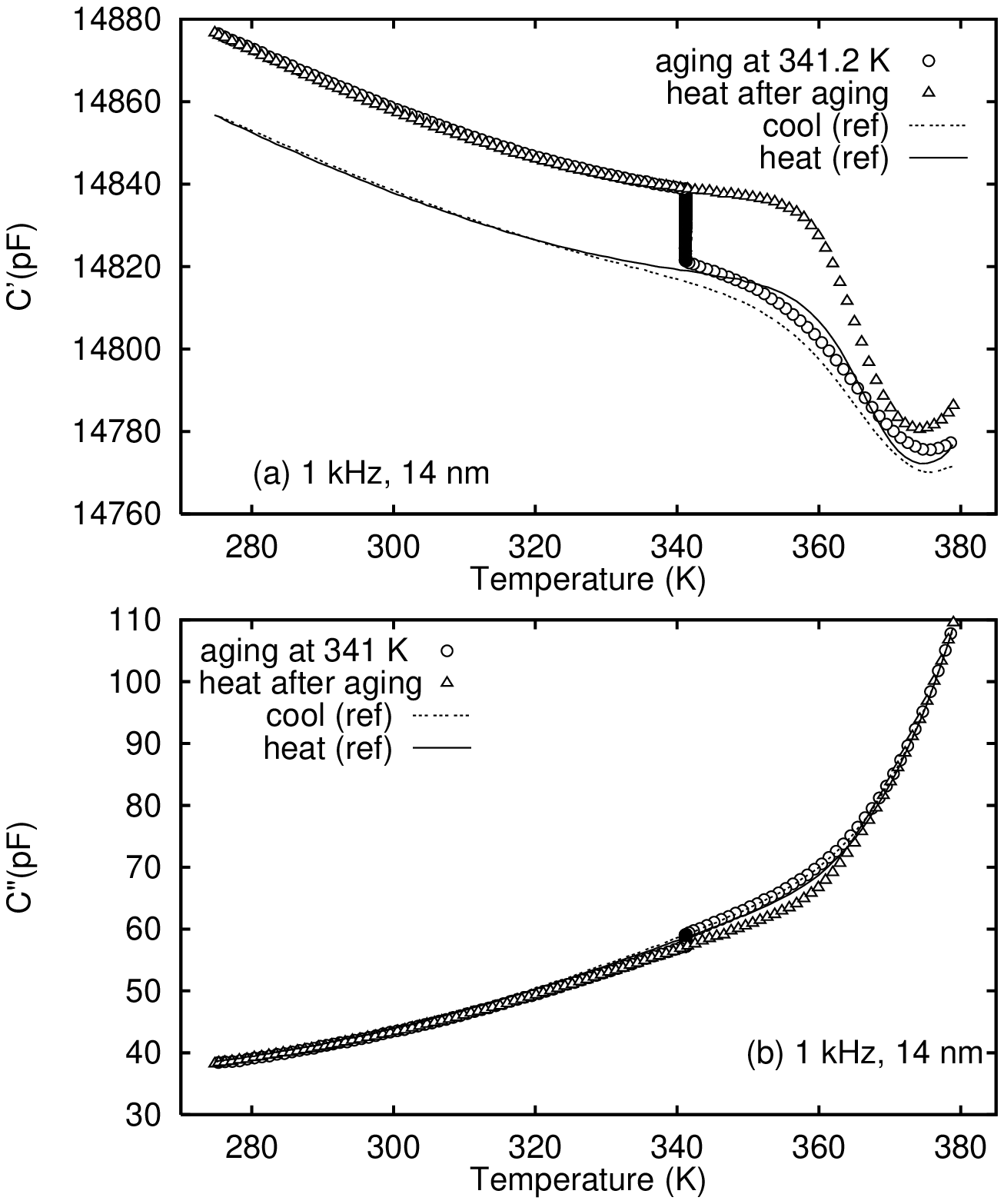}% Here is how to import EPS art
\caption{\label{fig:2} Temperature change in (a) the real part of electric
 capacitance $C'$ and (b) the imaginary part $C''$ for
 a-PS films with thickness of 14~nm for frequency 1~kHz. Temperature changes
between 380~K and 273~K including isothermal aging at 340~K. 
%(See the detailed explanation in text).
}
\end{figure}

Figure 2 shows the temperature dependence of the complex electric
capacitance in thin films of a-PS with $d=$~14~nm. In this case, a
similar result was obtained for the temperature dependence of $C'$ and $C''$
as that for $d=$~284~nm. 
For thin films, the absolute value of $C^*$ is enhanced.
The change in $C'$ is more distinct due to the smaller slope of $C'$
with respect to $T$. Furthermore, the scatter of data in $C'$ is 
reduced due to the larger geometrical capacitance.
The slight increase in $C'$ above 375 K was
due to the existence of the $\alpha$-process. 
In the case of $d=$~14~nm, the amount of the change in $C'$ and $C''$
are 0.1~\% and 3 ~\% for the isothermal aging at 340~K for 20 hours,
respectively.

This suggests that during the
isothermal aging process of a-PS, $C'$ increases with aging time,
whereas $C''$ decreases with aging time. It is important to note
that both $C'$ and $C''$ decrease with increasing aging 
time for the isothermal aging in 
poly(methyl methacrylate)~\cite{Bellon1,Fukao1,Fukao3}.
It is also reported in the literature that the real part of
complex electric capacitances (dielectric constants) 
decreases with increasing aging time for
polycarbonate~\cite{Cangialosi1} and poly(ethylene 
terephthalate)~\cite{McGonigle1}. The observed increase in $C'$ in 
a-PS is different from the decrease observed commonly for other polymeric 
systems.

\begin{figure}
%\vspace*{6cm}
%\includegraphics*[width=7.2cm]{./graph_ps025a_5f2_b.eps}
%\includegraphics*[width=7.2cm]{./graph_ps025a_5f2_a.eps}
\includegraphics*[width=7.2cm]{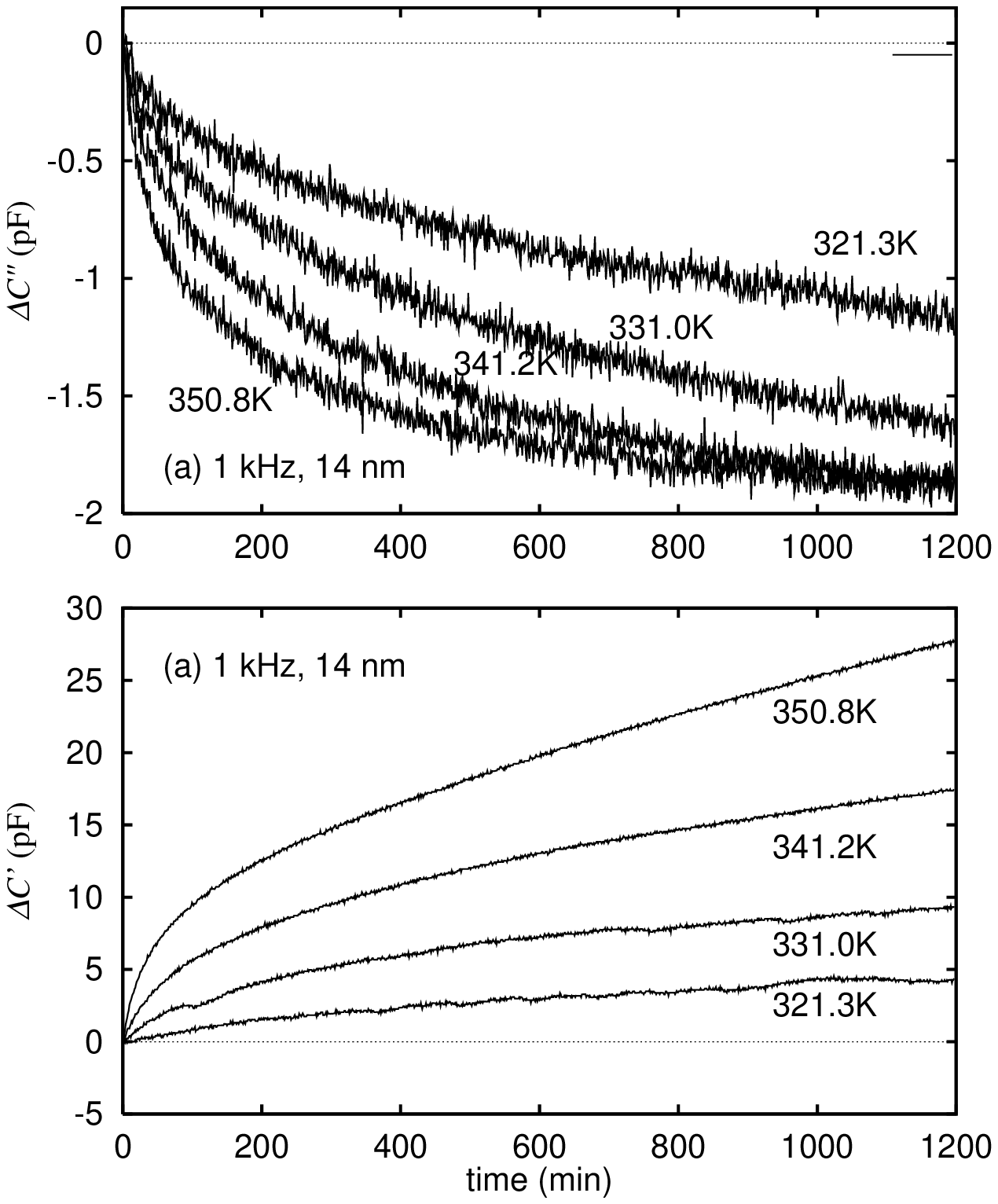}

\includegraphics*[width=7.2cm]{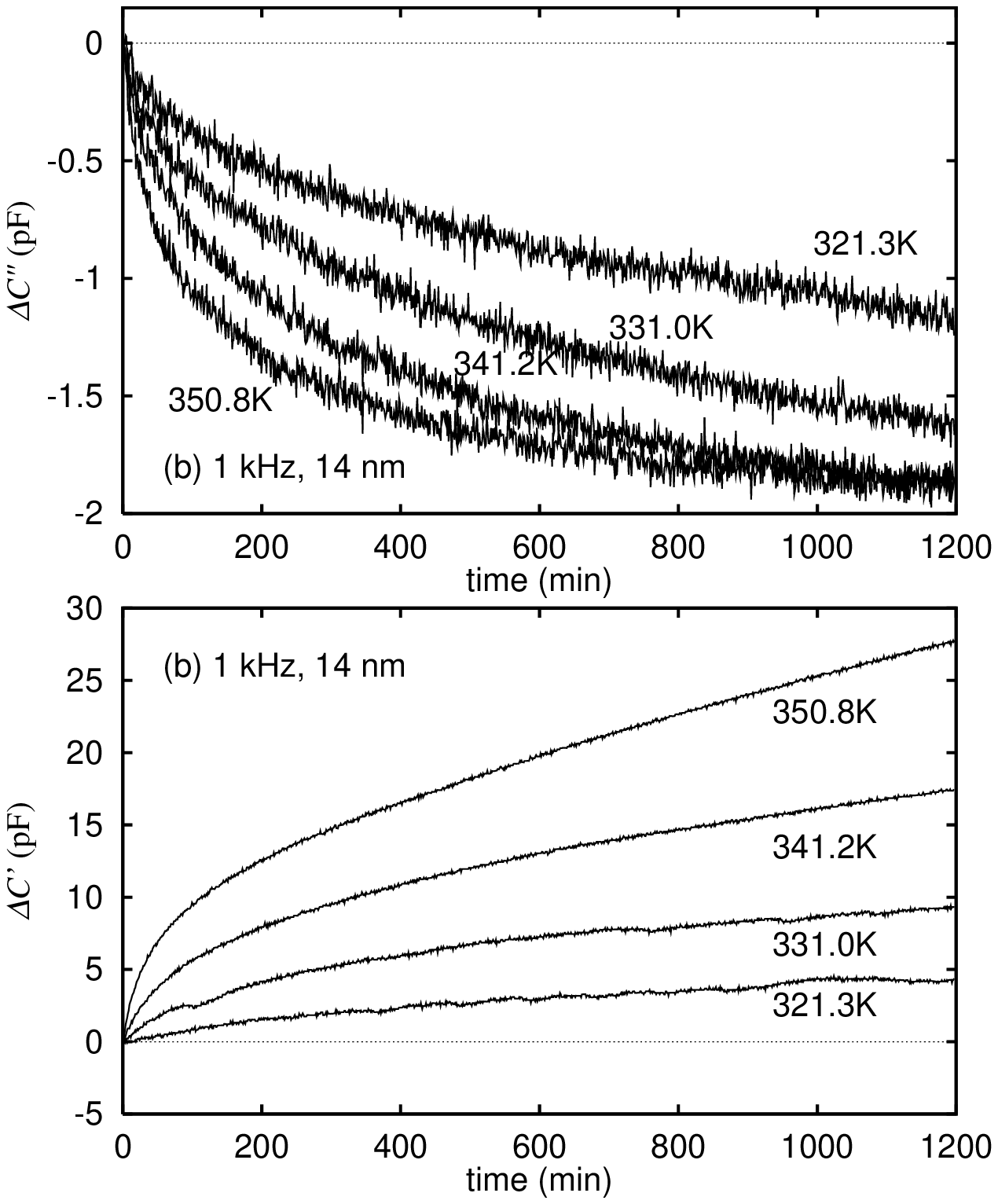}
% Here is how to import EPS art
\caption{\label{fig:3} Aging time dependence of the deviation 
$\Delta C'$ ($\Delta C''$) of $C'(t)$ 
($C''(t)$)  from the initial values $C'(0)$ ($C''(0)$) for various 
aging temperatures for thin films of a-PS with film thickness of 14~nm
 for $f$=1~kHz. The aging temperatures are 350.8~K, 341.2~K, 331.0~K and
 321.3~K.  
} 
\end{figure}

Figure 3 shows the aging time dependence of $C'$ and $C''$ during
isothermal aging at various temperatures $T_a$ (between 321.3 K and 350.8 K)
for $d=$~14~nm. 
The values $\Delta C'(t_{\rm w})$ and $\Delta C''(t_{\rm w})$ are
defined by $\Delta C'(t_{\rm w})$=$C'(t_{\rm w})-C'(0)$ and 
 $\Delta C''(t_{\rm w})$=$C''(t_{\rm w})-C''(0)$, and
the time $t_{\rm w}$=0 is the initial time at which the temperature 
of the sample reaches the aging temperature.
In Fig.~3(a), the value $\Delta C'$ 
increases monotonically with aging time and the 
value $\Delta C'$ at 20 hours
decreases with decreasing aging temperature. In Fig.~3(b), it is found
that $\Delta C''$ decreases with increasing aging time 
and the absolute value of $\Delta C''$ at 20 hours
decreases
with decreasing $T_a$ between 341.2~K and 321.3~K. The relaxation
behavior of $C''$ for isothermal aging at $T_a$=350.8 K appears to 
be slightly different from that at other $T_a$s lower than 350.8 K.

\subsection{Change in volume and dielectric permittivity}

A possible explanation for the observed dependence of $C'$ and
$C''$ on the aging time during isothermal aging is provided in 
this section. The real and imaginary parts of the complex electric
capacitance are given as follows:
\begin{eqnarray}
\label{Eq1}
C'(\omega,T,t)&=&(\epsilon_\infty (T,t)+\epsilon'_{\rm
 disp}(\omega,T,t))C_0(T,t)\\
\label{Eq2}
C''(\omega,T,t)&=&\epsilon''_{\rm disp}(\omega,T,t)C_0(T,t),
\end{eqnarray}
where $\epsilon_\infty$ is the dielectric constant at the high frequency
limit,and $\epsilon'_{\rm disp}$ and $\epsilon''_{\rm disp}$ are
frequency-dependent
contributions to the dielectric constant due to the orientational polarization
associated with molecular motions. The following relations can be derived:
$\epsilon'=\epsilon'_{\rm disp}+\epsilon_{\infty}$ and
$\epsilon''=\epsilon''_{\rm disp}$.  
In the case of a-PS,  the polarity is very weak, therefore; 
$\epsilon'_{\rm disp}\ll\epsilon_{\infty}$ is expected. Thus, 
Eq.~(\ref{Eq1}) can be approximately rewritten as 
\begin{eqnarray}
C'(\omega,T,t)&\approx&\epsilon_\infty(T,t) C_0(T,t).
\end{eqnarray}
For an isothermal aging process, the density is expected to increase with
aging time, and hence the film thickness $d$ decreases on the condition 
that the
area of the film remains constant. Therefore, this density change causes an
increase in $C'$ according to the following rate
\begin{eqnarray}\label{Eq4}
\frac{1}{C'}\frac{\partial C'}{\partial t}=-(\eta
 +1)\frac{1}{d}\frac{\partial d}{\partial t},
\end{eqnarray}
where $\eta$ is a constant and its value is almost equal to 1 for
 a-PS~\cite{Fukao1}. Eq.~(\ref{Eq4}) suggests that the decrease in film
 thickness increases $C'$ for isothermal aging.

\begin{figure}
%\vspace*{6cm}
%\includegraphics*[width=7.2cm]{./c_inv-t-1.eps}
\includegraphics*[width=7.2cm]{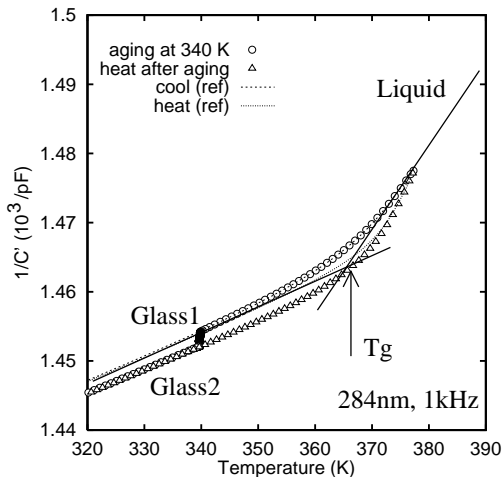}
% Here is how to import EPS art
\caption{\label{fig:4} Temperature dependence of the inverse of $C'$ for
 a constant rate mode observed at 1 kHz for a-PS with thickness of
 284~nm. The vertical axis approximately corresponds to the volume $V$.
} 
\end{figure}

Figure 4 shows the temperature change in the inverse of $C'$ observed
during the constant rate mode. The vertical axis approximately
corresponds to the volume of this system if the area remains constant. 
Therefore, in Fig.~4 
a temperature dependence of the volume characteristic of  
amorphous polymers can be observed. As the sample is cooled down 
from a high temperature
to a lower one through $T_{\rm g}$, the initial liquid state 
changes to a glassy state (Glass 1) and then the glassy state is converted
into a second glassy state (Glass 2) during an isothermal aging. 
This temperature dependence of
volume can typically be measured using dilatometric measurements
for amorphous polymers.

On the other hand, Eq.~(\ref{Eq2}) shows that $C''$ includes two different
 contributions from $\epsilon''_{\rm disp}$ and $C_0$. For the isothermal
 aging, $C_0$ is expected to increase with aging time, because the 
density increases,
 $i.e.$, the film thickness decreases, as shown above. 
Therefore, the decrease in $C''$ with aging time suggests that 
the imaginary part of dielectric constant, $\epsilon''_{\rm disp}$, 
decreases with aging time and that its contribution can overcome 
the contribution from the increase in $C_0$.
For PMMA, which has a strong polar group in the chain, it has been
 reported that both real and imaginary parts of the dielectric constant
 decrease with aging time during the isothermal aging 
process~\cite{Bellon1,Fukao5}.

The
results observed in a-PS can be explained as follows: for the
isothermal aging process, {\it the decrease in film thickness} is observed as
the increase in $C'$, whereas {\it the decrease in the imaginary component of
dielectric permittivity} is observed as a decrease in $C''$. Therefore, the
present measurement on the electric capacitance of a-PS will provide 
information
on the change in volume and dielectric permittivity simultaneously
for the same sample during the isothermal aging process.

\subsection{Volume relaxation during  isothermal aging}

\begin{figure}
%\vspace*{6cm}
%\includegraphics*[width=8.5cm]{./graph_ps3a_4ddx.eps}
\includegraphics*[width=8.5cm]{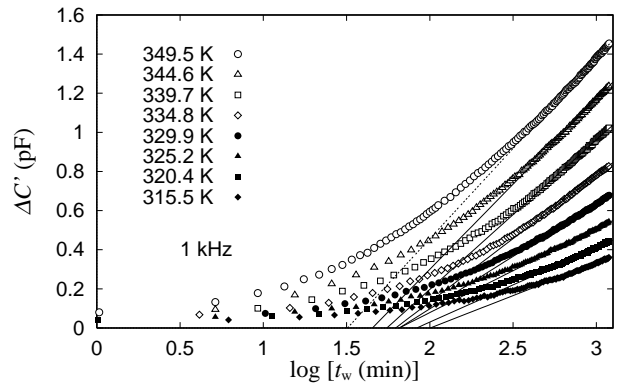}
% Here is how to import EPS art
\caption{\label{fig:5} Time evolution of the deviation $\Delta C'$ 
observed at various aging temperatures for a-PS thin films with
 $d=$ 284~nm.
The frequency of the applied electric field is 1 kHz. The horizontal
 axis is the logarithm of aging time.
} 
\end{figure}

As discussed in the previous section, the real part of the complex
electric capacitance changes with aging time in accordance with the
volume change. Figure 5 shows the aging time dependence of $C'$ relative
to the value of $C'(0)$ for thin films with $d=$~284~nm. The
horizontal axis is the logarithm of aging time. Figure 5 clearly shows
that $\Delta C'$ %C'-C'_{\rm ref}$ 
increases with the logarithmic law
with respect to the aging time for the long time region. Further, there is no
tendency to saturate in this time region. We evaluate the slope of this
aging time dependence using the following equation:
\begin{eqnarray}
\Delta C'(t_{\rm w}) &=& \tilde A\log t_{\rm w} + B \quad 
\mbox{(for large $t_{\rm w}$)}
\end{eqnarray}
Here, from the value of $\tilde A$, the aging rate of the volume, $A$,
for thin films of a-PS can be evaluated as follows:
\begin{eqnarray}
A&=&\frac{\tilde A}{C'(0)}
=\frac{1}{C'(0)}\frac{d\Delta C'}{d\log t_{\rm w}},
\end{eqnarray}
where for a given frequency $C'(0)$ is the value of $C'$ 
at the time at which the temperature reaches the aging temperature.
Figure~6 shows the temperature dependence of $A$ for thin films with
two different thicknesses $d=$ 14~nm and 284~nm. 
In the present temperature range, the aging rate $A$
was found to be larger in $d=$~284~nm than in $d=$~14~nm.
In this figure, it is
clear that the rate $A$ increases with increasing aging temperature and
obeys the Arrhenius law: $A=A_0\exp(-U/k_BT)$, where $k_B$ is the
Boltzmann constant, $U$ is the activation energy and $A_0$ is a constant.

\begin{figure}
%\vspace*{6cm}
%\includegraphics*[width=8.5cm]{./graph_slope_4.eps}
\includegraphics*[width=8.5cm]{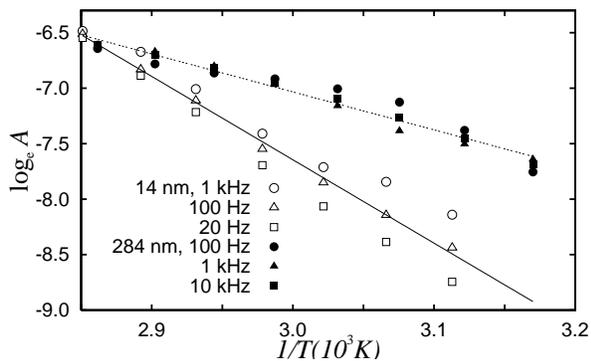}
% Here is how to import EPS art
\caption{\label{fig:6} Dependence of the logarithm of the coefficient $A$ 
on $1/T$ for various frequencies and thicknesses. 
The values of $f$ are 20 Hz, 100 Hz, and 1 kHz, and the values of $d$ are
14 nm and 284 nm.} 
\end{figure}

It should be discussed what is the molecular origin associated
with the volume change during the aging process. 
%The above analysis 
%revealed that the activation energy related to this change range 
%from 6.7~kcal/mol to 14.9~kcal/mol. 
Below $T_g$ there should be 
no contribution due to the $\alpha$-process, and hence it can
be expected that the $\beta$-process should be the most probable
candidate for the molecular origin for the physical aging.
However, in the literature 
the activation energy for the $\beta$-process in atactic 
polystyrene was estimated to be 38~kcal/mol~\cite{McCrum1} or 
30~kcal/mol~\cite{Yano1}, which are larger than the activation
energy evaluated from the temperature dependence of the aging rate $A$.
Therefore, the present result cannot be regarded as an 
evidence that the $\beta$-process is directly associated to the 
physical aging, although the $\alpha$-process can be excluded from
the candidate for molecular origin for the physical aging.
Probably, the $\beta$-process is indirectly related to the volume 
change during the physical aging process.

\subsection{Memory and rejuvenation in thin films}

\begin{figure*}

\vspace*{-0.2cm}
%\includegraphics*[width=7.5cm]{./graph_ps3a_4f2_1xy.eps}% Here is how to
 % import EPS art
%\includegraphics*[width=7.5cm]{./graph_ps025a_5f3a_ax.eps}% Here is how to
\includegraphics*[width=7.5cm]{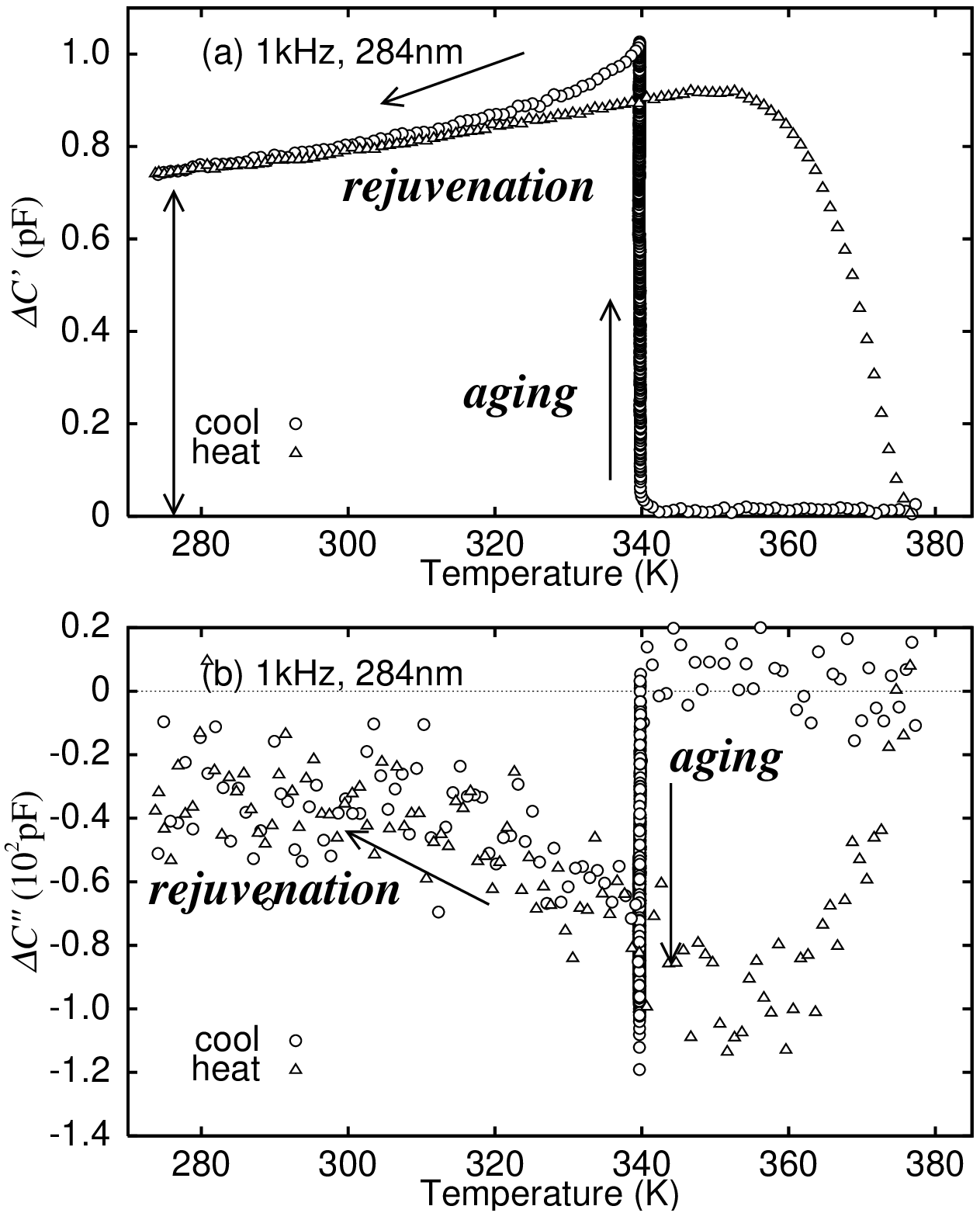}% Here is how to
 % import EPS art
\includegraphics*[width=7.5cm]{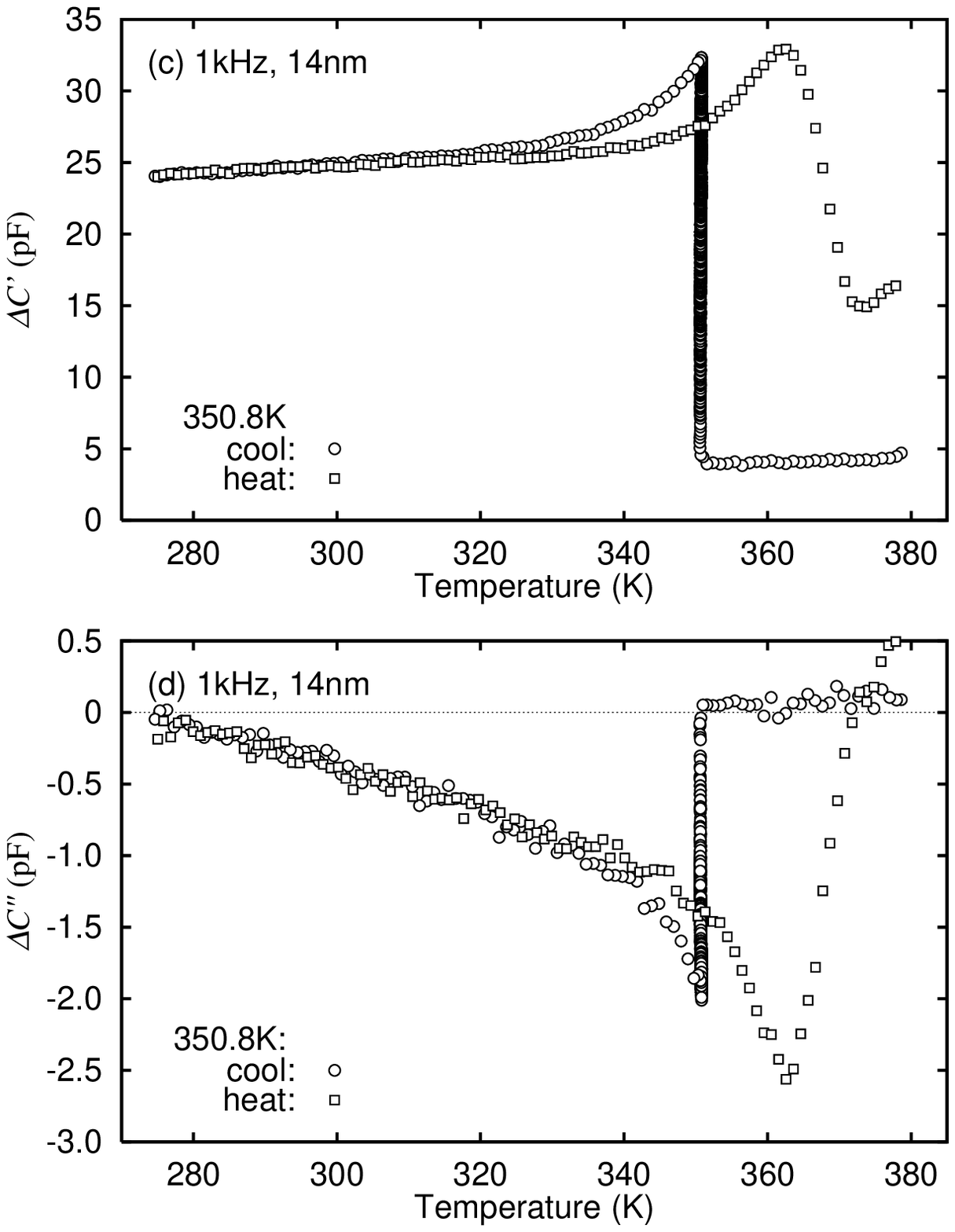}% Here is how to

\caption{\label{fig:7} 
Temperature dependence of the deviation $\Delta C'$ ($\Delta C''$) of 
the component of the complex electric capacitance $C'$ ($C''$) from the 
reference values observed for the cooling process including at an
 isothermal aging at 350.8~K and the subsequent heating process for thin
 films of a-PS with $d$~=~14~nm and $f$~=~1~kHz ((a) and (b))
and 100~Hz ((c) and (d)).} 
\end{figure*}

Figure 7 shows the temperature dependence of $\Delta C'$ and $\Delta C''$
observed during the cooling process with isothermal aging at $T_a$
and the subsequent heating process for 
$f$=1~kHz in a-PS films with $d=$~284~nm ((a), (b)) and 14~nm ((c),
(d)).
 In this case, $\Delta C'(T)$
($\Delta C''(T)$) is evaluated as the deviation of $C'(T)$ ($C''(T)$)
from the reference values $C'_{\rm ref}(T)$ ($C''_{\rm ref}(T))$ 
at the same temperature $T$. For the
reference value $C'_{\rm ref}$ ($C''_{\rm ref}$) to the cooling process 
including isothermal 
aging, we used the data measured for the preceding cooling process
without any isothermal aging, and for the reference value for
the heating process after the isothermal aging, we used the data
measured for the preceding heating process.

Figure 7(a) shows that as the temperature decreases from 380~K to 340~K,
$\Delta C'$ remains almost zero, and 
the deviation $\Delta C'$ increases as the aging time increases 
during isothermal aging at 340~K (See open circles). 
During the cooling process, after the isothermal aging, the deviation
decreases and then approaches a constant value, but {\it not 
zero}. As a result, most of the
deviation $\Delta C'$ induced during the isothermal aging remains even
at 273 K. This result is thought to be associated with the fact that the
isothermal aging increases the density.
For the subsequent heating process, $\Delta C'$ changes
along the path traced by $\Delta C'$ for the preceding cooling
process after isothermal aging, and then $\Delta C'$
reaches a maximum just above $T_a$ (See open triangles). 
The value of $\Delta C'$ subsequently decreases approaching zero. 
This behavior can be interpreted as follows: the fact that the sample
experiences aging at $T_a$ by the way of the preceding cooling process is
memorized within the sample, and the memory is recalled during the
subsequent heating process. 

The temperature dependence of $\Delta C''$ is
different from that of $\Delta C'$, as shown in Fig.~7(b).
For the cooling process from 380 K, $\Delta C''$ remains almost zero,
and then $\Delta C''$ decreases with increasing aging time for the
isothermal aging. $\Delta C''$ subsequently increases with decreasing
temperature and reaches zero at 273 K. 
In this case, there is a large scatter in the values of $C'$, because
the thickness is 284~nm and the geometrical capacitance $C_0$ is not
large enough. 
Figure 7(d) shows the temperature dependence of $\Delta C''$ 
during the constant rate mode for $d=$ 14~nm. 
For $d=$ 14~nm, the scatter of $C''$ is suppressed, therefore; it is 
clear that the value of $\Delta C''$ reaches zero at 273~K after cooling 
from $T_a$. This result suggests that the
system is {\it rejuvenated} as far as the dielectric response is
concerned. For the subsequent heating process, $\Delta C''$ decreases
along the curve observed during the preceding cooling process, and 
reaches zero after showing a minimum just above $T_a$.
Because the temperature dependence of $\Delta C''$ is primarily
attributed to that
of $\epsilon''$, it follows that $\epsilon''$ exhibits a 
{\it memory and rejuvenation effect}. 
A similar effect has been observed for $\epsilon'$
and $\epsilon''$ in the case of PMMA~\cite{Bellon1,Fukao5}.

Combining the results observed for $\Delta C'$ and $\Delta C''$, it is
concluded that the volume of a-PS films decrease during isothermal 
aging, and the deviation  from the reference value is
maintained below the aging temperature. On the other hand, dielectric
permittivity also decreases during isothermal aging, and the
deviation of the dielectric permittivity from the reference value 
is totally rejuvenated at lower temperatures. 
The existence of the volume change observed in
the present measurement is consistent with the fact that there are
several reports relating to the change in volume or density due to 
physical aging~\cite{Struick}. 
Based on the volume, the system does not appear to be rejuvenated, 
however, the dielectric response to the 
ac-electric field is fully rejuvenated.

In disordered systems, similar memory and rejuvenation effects are often
observed during the aging process. In spin glasses, magnetic
ac-susceptibility shows a memory and rejuvenation effect during the
constant rate mode and the temperature cycling mode. A possible
mechanism for this effect has been extensively discussed by a 
number of research
groups~\cite{Bray1,Fisher1,Fisher2,Berthier1,Jonsson1,Bouchaud1,Takayama1}.

In spin glasses, memory and rejuvenation effects have been investigated 
through the measurements of magnetic ac-susceptibility.
A number of experiments clearly show the existence of a strong rejuvenation
effect in aging phenomena through the so-called temperature-cycling
mode, where the aging temperature is maintained at $T_1$ for the first and 
third stage and at $T_2$ (smaller than $T_1$) for the second stage.
At the beginning of the second stage of the temperature-cycling
mode, the imaginary part of the magnetic ac-susceptibility 
$\chi''_{ac}$ returns to the initial value for $\chi''_{ac}$ at the
beginning of the first stage, although $\chi''_{ac}$ decreases during the
first stage. However, it should be noted here that the
magnetic ac-susceptibility is associated with a time scale,
defined by $t_{\omega}=2\pi/\omega$ (where $\omega$ is the angular frequency of
the ac-magnetic field). There is a relationship between the time
scale and the length scale~\cite{Bouchaud1}, therefore, the magnetic 
ac-susceptibility is
associated with the length scale $\xi(t_{\omega})$.
The experimental results suggest that the dielectric ac-susceptibility 
shows a rejuvenation effect, whereas the volume of the system changes with 
increasing aging time and shows no rejuvenation.
Although these findings appear to be inconsistent, this can be explained
as follows.
Similar to magnetic ac-susceptibility, the dielectric
ac-susceptibility is associated with a length scale $\xi(t_{\omega})$ 
defined by the angular frequency $\omega$ of the applied 
electric field. The volume or density is a macroscopic physical quantity
and has a characteristic length scale significantly larger 
than $\xi(t_{\omega})$
for the present frequency range. This suggests that, even if the volume changes
during the isothermal aging process and the change is maintained at
lower temperatures, the dielectric response to the applied electric field 
is not influenced by the change in volume. Therefore, the dielectric
ac-susceptibility can be rejuvenated during the cooling process after the
isothermal aging process.

If both magnetic ac-susceptibility and a macroscopic physical
quantity such as the remaining magnetization are measured 
simultaneously during the isothermal aging process, it can be
expected that for spin glasses the magnetic ac-susceptibility will show 
a rejuvenation, whereas the magnetization is not rejuvenated.\\

\section{Summary}

Herein, we investigated the aging dynamics for thin films of 
atactic polystyrene through measurements of complex electric 
capacitance using dielectric relaxation spectroscopy. 
The results obtained in this study are summarized as follows:\\

\begin{enumerate}
\item
During the isothermal aging process at a given aging temperature 
the real part of the electric capacitance increases
 with aging time, while the imaginary part
decreases with aging time.
\item
The aging time dependences of the real and imaginary parts of the
 electric capacitance are
 mainly associated with the change in  volume or film thickness and
 dielectric permittivity, respectively. 
\item
Memory effect can be observed for both ac-dielectric permittivity
and volume. On the other hand, a strong rejuvenation effect can
be observed for the ac-dielectric permittivity, but not for the volume.
\end{enumerate}

%\Appendix

\section{Acknowledgments}

The authors thank K.~Takegawa and Y.~Saruyama for useful
discussions.
This work was supported by a Grant-in-Aid for Scientific Research
(B) (No.~16340122) from the Japan Society for the Promotion of Science and
for Exploratory Research (No.~19654064) and Scientific Research in 
 the Priority Area ``Soft Matter Physics'' 
from the Ministry of Education, Culture, Sports, Science and Technology 
of Japan.
%%-----------------------------
%%      your bibliography
%%-----------------------------

%\newpage
\end{document}